\documentclass[pra,twocolumn,showpacs,preprintnumbers,amsmath,amssymb,superscriptaddress]{revtex4}
\usepackage{graphicx}
\usepackage{dcolumn}
\usepackage{bm}
\usepackage{amsmath}	
\usepackage{upgreek}
\begin{document}

\title{Integrated Atom Detector Based on \\Field Ionization near Carbon Nanotubes}

\author{B. Gr\"uner}
\author{M. Jag}
\author{A. Stibor}
\author{G. Visanescu}
\author{M. H\"affner}
\author{D. Kern}
\author{A. G\"unther}
\email{aguenth@pit.physik.uni-tuebingen.de}
\author{J. Fort\'agh}

\affiliation{CQ Center for Collective Quantum Phenomena and their Applications, Eberhard-Karls-Universit\"at T\"ubingen,
Auf der Morgenstelle 14, D-72076 T\"ubingen, Germany}

\date{\today}

\begin{abstract}
We demonstrate an atom detector based on field ionization and subsequent ion counting. We make use of field enhancement near tips of carbon nanotubes to reach extreme electrostatic field values of up to $9\times 10^9$V/m, which ionize ground state rubidium atoms. The detector is based on a carpet of multiwall carbon nanotubes grown on a substrate and used for field ionization, and a channel electron multiplier used for ion counting. We measure the field enhancement at the tips of carbon nanotubes by field emission of electrons. We demonstrate the operation of the field ionization detector by counting atoms from a thermal beam of a rubidium dispenser source. By measuring the ionization rate of rubidium as a function of the applied detector voltage we identify the field ionization distance, which is below a few tens of nanometers in front of nanotube tips. We deduce from the experimental data that field ionization of rubidium near nanotube tips takes place on a time scale faster than $10^{-10}$s. This property is particularly interesting for the development of fast atom detectors suitable for measuring correlations in ultracold quantum gases. We also describe an application of the detector as partial pressure gauge.

\end{abstract}
\pacs{79.70.+q, 81.07.De, 47.80.Fg}
\maketitle


\section{Introduction}

Probing ultracold quantum gases has been a driving force for the development of atom detectors. Absorptive, dispersive, and fluorescence imaging are routinely used to measure the density and momentum distribution of atom clouds \cite{Ketterle1999}. In combination with cavities, optical diagnostics reaches single atom sensitivity and allow measuring number statistics and correlations in quantum gases \cite{Mabuchi1996,Muenstermann1999,oettl2005,Teper2006,Colombe2007}. Alternative detection techniques have been developed based on electron impact ionization \cite{Gericke2008} or photoionization of atoms \cite{Guenther2009} with subsequent ion counting. Direct counting of metastable atoms on a microchannel plate has been demonstrated to give high spatial and temporal resolution \cite{Robert2001}, properties which are desirable also in experiments with ground state atoms and molecules. In this manuscript, we describe a detector based on field ionization of ground state atoms near tips of carbon nanotubes and subsequent ion counting. This neutral particle detection scheme does not require optical fields and has the potential to reach high spatial and temporal resolution of detection. In addition, it can be ideally integrated to atom chips, where a localized probe tip or an array of ionizing nanotubes may be used to detect atoms and molecules in-situ or in time-of-flight.

\begin{figure}
        \centerline{\scalebox{0.7}{\includegraphics{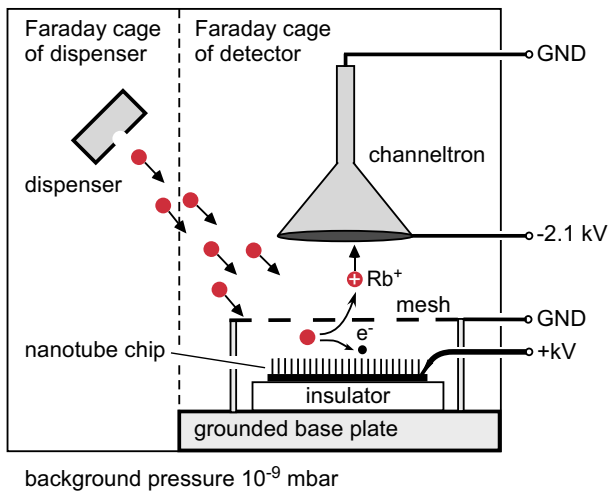}}}		
		\caption{(Color online) Atom detector based on field ionization and subsequent ion counting (not to scale). A two dimensional array of multiwall carbon nanotubes on a substrate (size of $10 \mbox{mm} \times 20 \mbox{mm}$) defines the ionization area of the detector. The nanotube chip is on a positive potential below a grounded mesh. Atoms entering the detector are field ionized near nanotube tips and guided to a channel electron multiplier for detection. The setup is placed in a Faraday cage in order to be insensitive to external stray fields. We demonstrate the operation of the detector by counting atoms from a thermal beam of a rubidium dispenser. The dispenser is in a separate faraday cage next to the detector and is on a negative potential above ground in order to stop direct emission of ions. The detector is operated in ultrahigh vacuum at $\sim 10^{-9}$ mbar.}
	\label{fig:Scheme1}
\end{figure}

Electric fields that ionize ground state alkali atoms are on the order of $10^9-10^{10}$ V/m \cite{Beckey1977} and are hardly achievable in free space. We make use of the enhancement of the electric field near conducting tips \cite{Jackson1998} to reach these extreme values locally. We place carbon nanotubes (CNTs) into an electrostatic field parallel to their axis. Due to the high aspect ratio of nanotubes, electric charges concentrate on the tube tips resulting in a field enhancement of up to several orders of magnitude. Our nanotubes are grown on a substrate perpendicular to the surface (Fig. \ref{fig:Scheme1}). The substrate and a metal mesh mounted above it serve as electrodes to set the voltage that defines the electrostatic field along the nanotubes. We reach the ionizing field already at moderate voltages. Atoms hitting the field enhanced area in the vicinity of nanotube tips are ionized and repelled from the positively charged surface. The ions are guided to a channel electron multiplier (CEM) above the mesh and are detected with single ion resolution (Fig. \ref{fig:Scheme1}).

We demonstrate the nanotube ionization detector in ultrahigh vacuum by counting rubidium atoms from a thermal beam of a dispenser source. We quantify the electric field enhancement near nanotube tips by measuring the field emission of electrons. We prove that field ionization of rubidium takes place at distances below a few tens of nanometers in front of nanotube tips. We measure ionization rates of rubidium atoms above a nanotube array as a function of the applied voltage on the electrodes and as a function of the rubidium flux. We conclude that besides recently reported applications in gas analysis \cite{Modi2003,Riley2003}, carbon nanotube ionization detectors are suitable detectors for cold atom experiments and as partial pressure gauge at ultrahigh vacuum level.

\section{Setup of the field ionization detector}

\begin{figure}[tb]
\centerline{\scalebox{0.55}{\includegraphics{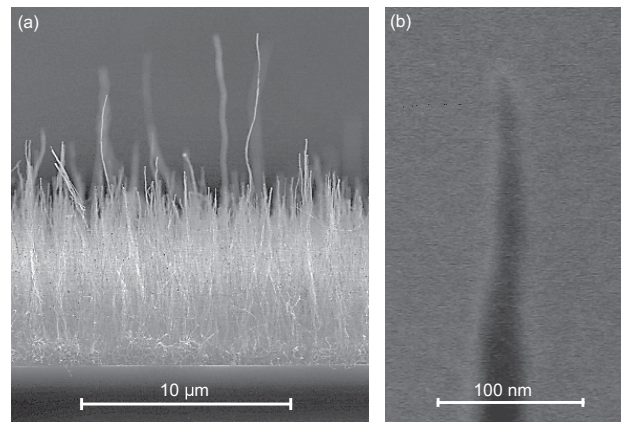}}}
\caption{Scanning electron microscope (SEM) image of the nanotube chip. (a) Side view of the carpet of multiwall carbon nanotubes on the silicon substrate. The typical length and diameter of the CNTs are $(8.6 \pm 1.1) \mu$m and $(58 \pm 11)$nm, respectively. The average distance between nanotubes is $100$nm. Individual nanotubes are significantly higher, up to $\sim 20 \mu$m, and exhibit sharp tips. (b) The tip of a $17.4 \mu$m long nanotube has a diameter of less than $15$nm (resolution limit of SEM), which is smaller than the average width of the nanotube $\sim 60$nm, giving an aspect ratio more than 1200.}
\label{fig:CNT}
\end{figure}

The central part of the field ionization detector is a carpet of multiwall carbon nanotubes that is grown on top of a silicon substrate (Fig. \ref{fig:CNT}). The $250 \mu$m thick substrate has a size of $10 \mbox{mm} \times 20 \mbox{mm}$, it is covered by a $7.5 \mbox{nm}$ thick silicon oxide layer and a $3 \mbox{nm}$ thick nickel layer on top of the oxide \cite{processing2009}. The nanotubes have an average length of $l=(8.6 \pm 1.1) \mu$m and a typical diameter of $\rho=(58\pm 11) $nm, resulting in a typical aspect ratio of $\eta=l/\rho=150$. The inter-tube distance is about $100 \mbox{nm}$. Individual nanotubes of the carpet are significantly longer and exhibit sharp tips (Fig. \ref{fig:CNT}(b)). We find nanotubes with aspect ratios of up to several thousand (cf. cylindrical nanotube model in Section III. A).

The nanotube chip is mounted on a 1mm thick glass plate in order to electrically isolate it from the grounded base plate (Fig. \ref{fig:Scheme1}). A stainless steel mesh is mounted at $d=1.25 \mbox{mm}$ above the chip surface. The mesh is made of $110 \mu$m thick wires with $508 \mu$m spacing, resulting in a transmission above $60 \%$ \cite{TWP}. The nanotube chip and the mesh are electrically contacted. Applying a voltage $U$ between the nanotube chip and the grounded mesh defines a homogeneous electric field according to the equation of a plate capacitor $F_{c}=U/d$. However, the local field $F$ at the tips of the nanotubes is enhanced due to the increased charge density by a factor of $\gamma:=F/F_{c}=Fd/U$ \cite{Li06}. The field enhancement factor $\gamma$ depends on the geometry and distribution of nanotubes on the substrate and is discussed in Section III. A. By applying a negative voltage to the nanotube chip, the device operates as electron field emitter. We use this operational mode to characterize the field enhancement near nanotube tips. By applying a positive voltage to the nanotube chip, the nanotube tips are ionizing neutral atoms and accelerate derivative ions towards the mesh. We use this operational mode for atom detection.

A channel electron multiplier (CEM) is mounted 24mm above the nanotube chip to detect charged particles. Ions (electrons) passing through the mesh are attracted by the CEM, which is on a negative (positive) potential on the order of few kV. A charged particle hitting the CEM releases an avalanche of secondary electrons in the CEM, resulting in a charge pulse that is amplified \cite{F100TD} and detected with standard counting electronics \cite{SR400}. To keep ion (electron) trajectories uninfluenced from electric stray fields, the detector is placed in a Faraday cage.

We demonstrate the operation of the field ionization detector by counting rubidium atoms from a dispenser source \cite{SAES}. Such dispensers are widely used in cold atom experiments for loading magneto-optical traps, the first step in the preparation of quantum gases \cite{Fortagh1998b}. We characterize the constituents emitted by the rubidium dispenser in the Appendix of this article and find that the emission of constituents other than rubidium can be neglected. In addition, we find that the ionization energy of rubidium ($4.18 \mbox{eV}$ \cite{Steck2008}) is by more than a factor of three smaller than those of other constituents, thus in our setup rubidium is the most sensitive element to field ionization.

The dispenser is directed towards the ionization detector and heated resistively. The emission of rubidium atoms starts at a temperature of $\sim 650$ K, and the atoms, with a mean thermal velocity of $\sim 400 \mbox{m/s}$, impact the nanotube carpet directly. Since the dispenser not only emits atoms but also positively charged rubidium ions, we lift the dispenser to a negative potential of $-2V$ with respect to ground. This attractive potential is sufficient to remove ions from the thermal beam. An additional Faraday cage around the dispenser (Fig. \ref{fig:Scheme1}) makes sure that external fields are absent and all dispenser ions are attracted by the dispenser's body. Hence, we reduce the background ion counting rate measured with the CEM from $7500$/s to $1$/s, at a dispenser current of $5.5$ A. The experiments are done in ultrahigh vacuum at a background pressure of $\sim 10^{-9}$ mbar.


\section{Characterization of the field ionization detector}

\subsection{Field enhancement near tips of nanotubes}

The field enhancement at the tip of a free standing carbon nanotube can be estimated by modeling the tube as a conducting cylinder with a hemispherical tip. Within this model, the field enhancement factor $\gamma_\infty$ depends only on the aspect ratio $\eta$ (length over diameter) of the nanotube and is given by \cite{Forbes2003}:
\begin{equation}
	\gamma_\infty= 1.2 \left( 2.15 + \eta\right)^{0.9} \;\sim\; \eta.
	\label{forbes}
\end{equation}
In an array of nanotubes, such as the carpet of carbon nanotubes in our experiment, the electrostatic field of neighboring nanotubes superimposes and reduces the field enhancement. The field enhancement factor $\gamma$ thus depends not only on the aspect ratio of the constituting tubes, but also on the ratio of inter tube separation $h$ to the tube length $l$ \cite{Bonard2001}:
\begin{equation}
	\gamma = \gamma_\infty \left[ 1 - \exp\left(- 2.3172 \; \frac{h}{l}\right)\right].
	\label{bonard}
\end{equation}
Using this model, we expect for the majority of nanotubes (aspect ratio: $\sim 150$, inter tube separation: $\sim 100$nm) a field enhancement factor of $\gamma \sim 3$ ($\gamma_\infty \sim 110$). However, nanotubes sticking out of the carpet reach much higher factors. This is because their larger aspect ratio, on the one hand, and also because the influence of neighboring nanotubes can be neglected, on the other hand. Recognizing that individual nanotubes on our chip have aspect ratios of up to $1200$, field enhancement factors on the order of thousands are expected. By applying a voltage to the nanotube chip on the order of kilovolts, the electric field near individual nanotubes should reach the value of $\sim 10^9$ V/m, which is sufficient to field ionize ground state rubidium atoms \cite{Beckey1977}. We notice that in the range of kilovolts, we expect only the longest nanotubes to contribute to field ionization.

We measure the field enhancement factor on our nanotube chip by measuring the field emission of electrons, i.e. by measuring the electric current tunneling out of negatively charged nanotube tips. The current-field characteristic of metallic field emitters is given by the Fowler-Nordheim equation \cite{Fowler28},
\begin{equation}
	I = A a \frac{F^2}{\varPhi}\; \exp\left(-b \frac{\varPhi^{3/2}}{F}\right) = RF^2\; \exp\left(-\frac{S}{F}\right)~,
	\label{eq:Fowler}
\end{equation}
with the emission current $I\left[\mbox{A}\right]$, the electric field $F\left[\mbox{V/m}\right]$, the emission area $A\left[\mbox{m}^2\right]$, and the metal's local work function $\varPhi\left[\mbox{eV}\right]$. The universal constants a and b are given by the electron's charge and mass, $e$ and $m_{e}$, and the Planck constant $\hbar = h/2\pi$:
\begin{eqnarray}
	a & = & \frac{e^3}{16\pi^2\hbar} = 1.5414 \times 10^{-6} \; \mbox{A}\;\mbox{eV}\;\mbox{V}^{\scriptscriptstyle{-2}} \label{eq:a}\\
    b & = & \frac{4\sqrt{2m_e}}{3e\hbar} = 6.8309 \times 10^9 \; \mbox{V}\;\mbox{m}\;\mbox{eV}^{\scriptscriptstyle{-3/2}}. \label{eq:b}
\end{eqnarray}
For simplicity, we introduced the system constants $R=Aa/\varPhi$ and $S=b\varPhi^{\scriptscriptstyle{3/2}}$ on the right hand side of Eq. \ref{eq:Fowler}. Since we are interested in the current-voltage characteristic, we rewrite the electric field $F$ in terms of the voltage $U$ applied between the electrodes of the nanotube chip, and introduce $\beta:=F/U$. The Fowler-Nordheim equation reads in its semi-logarithmic form as
\begin{equation}
	\ln\left(\frac{I}{U^2}\right) = \ln\left(R\beta^2\right) - \frac{S}{\beta U}.
	\label{eq:Fowlerlog}
\end{equation}
The linear dependence of $\ln(I/U^2)$ on $1/U$ is a key feature of field emission from a conducting surface and gives a convenient protocol to evaluate experimental data. The slope $S/\beta$ of a measured $\ln(I/U^2)$ vs. $1/U$ diagram (Fowler-Nordheim plot) yields the field proportionality constant $\beta$ (if $\varPhi$ is known), from which the electric field at the emitting surface can be directly calculated $F=\beta U$. The intercept $\ln(R\beta^2)$ provides in addition information about the emission area $A$.

The above description relys on ideal conditions \cite{Forbes1999}, which are difficult to fulfil with carbon nanotubes. Although a more general Fowler-Nordheim theory has been developed in the same reference, a precise theory for field emission of electrons from carbon nanotubes is still not available. Hence, the above description is widely used in the nanotube community and is accepted as fair approximation. We evaluate our experimental data using Eq. \ref{eq:Fowler} - \ref{eq:Fowlerlog}.

\begin{figure}
\centerline{\scalebox{0.65}{\includegraphics{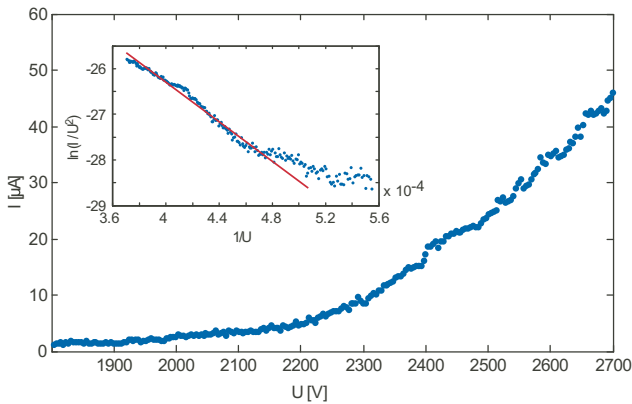}}}
\caption{(Color online) Field emission current of the nanotube chip as a function of the applied voltage. The inset shows the same data in the Fowler-Nordheim plot (semi-logarithmic $\ln(I/U^2)$ vs. $1/U$ diagram). The slope and intercept of the linear fit (straight red line) is used to estimate the field proportionality factor, $\beta\approx 3.3\times 10^6\mbox{m}^{-1}$, and the emission area, $A\approx 6.1\times 10^{-15}\mbox{m}^2$.}
	\label{fig:FieldEmission}
\end{figure}

In the experiment, we apply a variable negative voltage to the carbon nanotube chip and measure the field emission current using the amperemeter of our high voltage supply \cite{FUG}. In order to be well above noise level, we restrict our study to the voltage range between 1800 and 2700V. Figure \ref{fig:FieldEmission} shows the measured current-voltage characteristic and the corresponding Fowler-Nordheim plot.
The semi-logarithmic plot shows a linear behavior in the voltage range between 2100 and 2700 V. Fitting a straight line to this data, we extract the slope $S/\beta$ and the intercept $\ln(R\beta^2)$. Using the literature value of the nanotube's work function $\varPhi=4.8$eV \cite{Zhao2002}, we calculate the field proportionality factor, $\beta\approx 3.3\times 10^6\mbox{m}^{-1}$, and the emission area, $A\approx 6.1\times 10^{-15}\mbox{m}^2$. With $F=\beta U$ and $F=\gamma U/d$ we estimate the field enhancement factor, $\gamma=\beta d\sim 3900$, and the field strength, $F=\beta U\approx (6 - 9)\times 10^9$V/m, at the surface of the emitting nanotube tips.

From Eq. \ref{forbes} we estimate the aspect ratio of the emitting tubes to be $\eta\sim 8000$. Thus only the longest nanotubes with the sharpest tips of a few nm radius contribute to the signal. Only these are able to reach such a field enhancement. We estimate the number of nanotubes contributing to the field emission by representing the emission area $A_0$ of a single contributing nanotube as the surface of a hemisphere with radius 1nm. The number of contributing nanotubes is thus $A/A_0\sim 1000$. As the total number of nanotubes on the chip is about $10^{10}$, only one tube out of $10^7$ contributes to the field emission in the given voltage range. We conclude that the measured electric field at the surface of individual nanotubes on our chip, up to $9\times 10^9$V/m, is indeed sufficient to field ionize rubidium atoms.

\subsection{Field ionization of rubidium atoms near carbon nanotubes}

Field ionization of atoms in electrostatic fields is described by quantum mechanics as tunneling out of an electron from the distorted Coulomb potential into free space \cite{Beckey1977}. The ionization rate $\Gamma$ is associated with the tunneling rate and is given by
\begin{equation}
	\Gamma \sim \exp\left(-b\; \frac{W_{\mbox{\scriptsize ion}}^{3/2}}{F_i}\right) = \exp\left(-b\; \frac{W_{\mbox{\scriptsize ion}}^{3/2}}{\beta_i U}\right),
	\label{eq:ionization1}
\end{equation}
with $W_{\mbox{\scriptsize ion}}$ being the ionization energy of the atom, $F_i$ the electrostatic field in free space, and $b$ a natural constant given by Eq. \ref{eq:b}. We introduced again the field proportionality factor $\beta_i:=F_i/U$, since the electric field is controlled via a voltage $U$, which is applied between the electrodes of the nanotube detector. The index $i$ stands for ionization and is to distinguish between the ionizing field (with index $i$) and for the tubes surface field required for field emission of electrons (no index).
Equation \ref{eq:ionization1} is similar in form to the Fowler-Nordheim equation. The semi-logarithmic plot of the ionization rate vs. $1/U$ is again a straight line:
\begin{equation}
	\ln(\Gamma)=\ln C\;-\;\frac{b W_{\mbox{\scriptsize ion}}^{3/2}}{\beta_i}\,\times\,\frac{1}{U}.
	\label{eq:ionization2}
\end{equation}
Here, $C$ accounts for the proportionality in Eq. \ref{eq:ionization1} with a linear dependence on the atom flux. The equation gives a protocol to deduce the ionizing field from experimental data. If the ionization energy $W_{\mbox{\scriptsize ion}}$ is known, the slope of the semi-logarithmic plot gives $\beta_i$ and thus the field $F_i$ inside the ionization volume.

In the following, we describe two experiments that demonstrate the operation of the field ionization detector. Before the detector is first used, the nanotube chip is cleaned by exposing it to a high positive voltage (5000 V) for several minutes, which removes adsorbed atoms and molecules from the nanotubes. Subsequently, the voltage is reduced to the operating value of 3500 V or below. Once the CEM is activated by applying a negative voltage, -2.1 kV, to the front funnel (cf. Fig. \ref{fig:Scheme1}), the detector is ready for field ionization and ion counting. The rubidium flux entering the detector is activated and controlled by a heating current through the rubidium dispenser.

\begin{figure}
\centerline{\scalebox{0.65}{\includegraphics{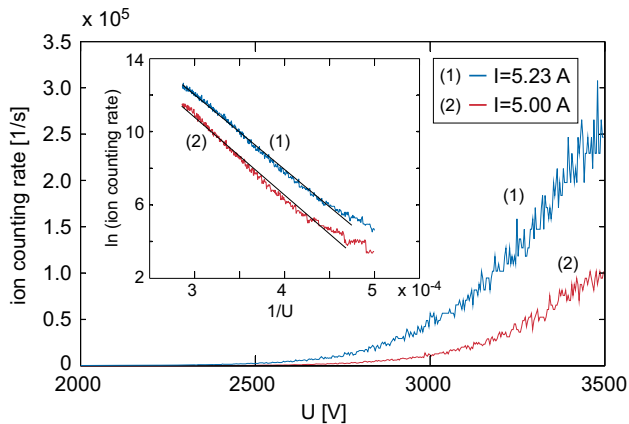}}}
	\caption{(Color online) Field ionization of rubidium atoms near carbon nanotubes. The diagram shows the ion counting rate at the CEM, proportional to the field ionization rate, as a function of the applied voltage on the nanotube chip. Measurements were taken for two different fluxes of rubidium atoms corresponding to dispenser currents of 5.23A [blue line (1)] and 5A [red line (2)]. The inset shows the semi-logarithmic Fowler-Nordheim equivalent plots of the same data. The linear behavior of the data gives evidence for field ionization of rubidium near nanotubes. By fitting the data using Eq. \ref{eq:ionization2} (straight black line) we find the value of the field proportionality factor $\beta_i=1.4\times 10^6\mbox{m}^{-1}$.}
	\label{fig:ionization1}
\end{figure}

In our first experiment we measure the field ionization rate as the function of the applied detector voltage. We use the data to identify the field ionization area near nanotube tips. At constant rubidium flux, we sweep the detector voltage linearly from 3500 to 2000 V within 40 s. The measured ion counting rate is plotted in Fig. \ref{fig:ionization1} for dispenser currents of 5.23 A (blue curve (1)) and 5.0 A (red curve (2)). The data are corrected for the background count rate that we measure under identical conditions but zero dispenser current. We notice that the rubidium flux for the two different heating currents differ by more than a factor of two. The Fowler-Nordheim equivalent plot in the inset in Fig. \ref{fig:ionization1} shows a linear behavior, which is characteristic for field ionization. By fitting the theory (Eq. \ref{eq:ionization2}) to the experimental data, we find the field proportionality factor, $\beta_i\approx 1.4\times 10^6\mbox{m}^{-1}$, which is the same within $5\%$ for both lines. As a result, the electric field at which rubidium atoms are ionized can be estimated to be $F_i\approx (3 - 4.5)\times 10^9\mbox{V/m}$.

Comparing the field proportionality factors measured by the field emission of electrons and by the field ionization of rubidium atoms, we realize that $\beta_i=0.42\beta$. Hence, the atoms are ionized before they reach a nanotube surface. The electric field ($F_i=\beta_i U$) at the position of field ionization is less than $50\%$ of the field at the tip surface ($F=\beta U = 2.36 F_i$). Because the electrostatic field near nanotube tips decays over a short distance, on the length scale of the tip radius ($1-40$nm) we estimate that the spatial area in which field ionization takes place is in our experiment below 40 nm distance in front of the nanotube tips. Based on the mean atomic velocity, $\sim 400$m/s, we calculate that the ionization takes place on a time scale faster than $10^{-10}$s. This finding is particularly interesting for the development of fast atom detectors for correlation measurements in cold gases.

\begin{figure}
	\centering
\centerline{\scalebox{0.53}{\includegraphics{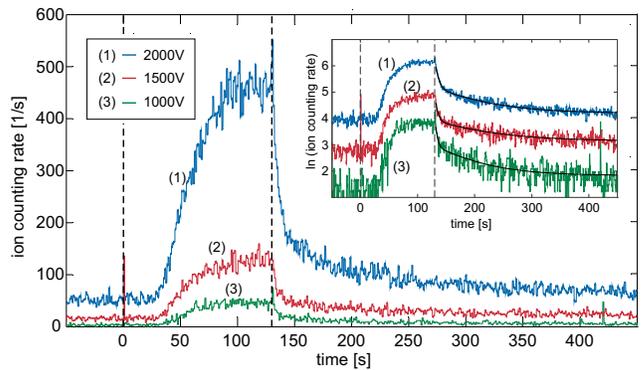}}}
	\caption{(Color online) Response of the field ionization detector to a time varying flux of rubidium atoms. The ionization rate is shown for three different voltages, 2000V, 1500V, and 1000V on the nanotube chip \cite{ionizationrate}. Heating of the rubidium dispenser with 5.9 A starts at $t=0$ (left dashed line) and stops at $t=130 s$ (right dashed line). Independent of the applied detector voltage, the ionization signal saturates at a factor of 10 above the background counting rate. The inset shows the semi-logarithmic plot of the same data to illustrate the double exponential decay after turning off the dispenser. The measured decay times are independent of the detector voltage, $\tau_1=(4.0 \pm 0.9) \mbox{s}$ and $\tau_2=(76 \pm 10) \mbox{s}$, and correspond to the reduction of the rubidium flux when the dispenser cools down and to the reduction of the rubidium background pressure determined by the pumping speed of the vacuum system, respectively.}
	\label{fig:OnOff1}
\end{figure}

In a second experiment we demonstrate the application of the field ionization detector as partial pressure gauge. We set the nanotube chip to a constant voltage and measure the response of the detector to a time varying flux of rubidium atoms. The rubidium flux is activated by switching on the heating current (5.9A) of the dispenser at $t=0$ and switching it off at $t=130 s$. The ion counting rate measured for various voltages at the nanotube chip, 2000 V, 1500 V, and 1000 V, is shown in Fig. \ref{fig:OnOff1}. In this voltage range the detector ionizes only rubidium (cf. Appendix and section above). During the heating of the dispenser, the ion count rate increases rapidly, it saturates on the time scale of about 100 s, and drops after the dispenser is switched off. The ion count rate after switching off is well described by a double exponential decay (inset in Fig. \ref{fig:OnOff1}). We attribute the fast timescale, $\tau_1=(4.0 \pm 0.9) \mbox{s}$, to the cooling down of the dispenser and a corresponding reduction of the rubidium flux. The data is in good agreement with previously reported values \cite{Fortagh1998b}. The slow time scale, $\tau_2=(76 \pm 10) \mbox{s}$, is due to rubidium background pressure that decays on a time scale given by the pumping speed of our vacuum system.

\begin{figure}
	\centering
\centerline{\scalebox{0.67}{\includegraphics{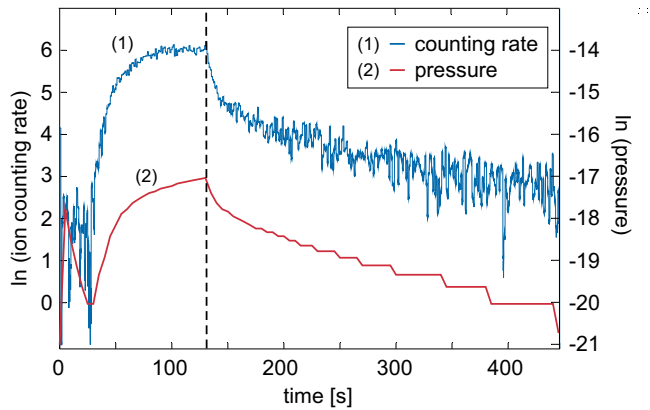}}}
			\caption{(Color online) Comparison of the signal of the field ionization detector [blue curve (1), legend left] and the pressure measured by a commercial pressure gauge [red curve (2), legend right] during a pulsed operation of the rubidium dispenser (semi-logarithmic plot). The pressure gauge detects any gas constituents while the field ionization detector is set to measure the rubidium only (see text). The diagram illustrates that after a dispenser pulse rubidium dominates the background pressure.}
	\label{fig:OnOff3}
\end{figure}

A comparison between the ion counting rate and the pressure measured simultaneously with a commercial pressure gauge \cite{Pfeiffer} confirms our conclusion (Fig. \ref{fig:OnOff3}) that the field ionization detector is suitable for measuring the rubidium partial pressure. The qualitative difference between the curves in Fig. \ref{fig:OnOff3} is the pressure peak between 0 and 30 s. This peak is due to a degassing of the dispenser shell when heating is turned on. The peak corresponds to vacuum background gas constituents mostly other than rubidium (cf. Appendix). The field ionization detector is not sensitive to these constituents since its discrimination voltage (operation voltage) has been set below their ionization limit. Thus the detector counts in this case only rubidium atoms.

\section{Conclusion}

The demonstrated field ionization detector outlines a range of new applications in cold atom physics. Since ground state atoms and molecules can be field ionized near nanotips, field ionization detectors are suitable for in-situ or time-of-flight measurements on cold gases without using resonant light and optical imaging systems. Field ionization detectors can be ideally integrated onto atom chips, where a single ionizing nanotube or an array of nanotubes may be used to measure spatial and temporal correlations in quantum gases \cite{Nara1999}. As field ionized atoms are detected by a channel electron multiplier, the detector intrinsically allows for detecting with single atom sensitivity. Due to this property, such a device may improve state of the art pressure measurements in terms of minimum detectable pressure.

The authors greatly acknowledge Ph. Schneewei{\ss}, M. Gierling, and P. Federsel for experimental support, and R. Forbes, M. Fleischer, C. Zimmermann, and T. E. Judd for valuable discussions. The work has been supported by the Bundesministerium f\"ur Bildung und Forschung (BMBF, NanoFutur 03X5506) and by the Deutsche Forschungsgemeinschaft (DFG, TRR21 Project C9).

\section{Appendix}

\subsection{Mass spectrum of a rubidium dispenser}
\begin{figure}[tb]
\centerline{\scalebox{0.65}{\includegraphics{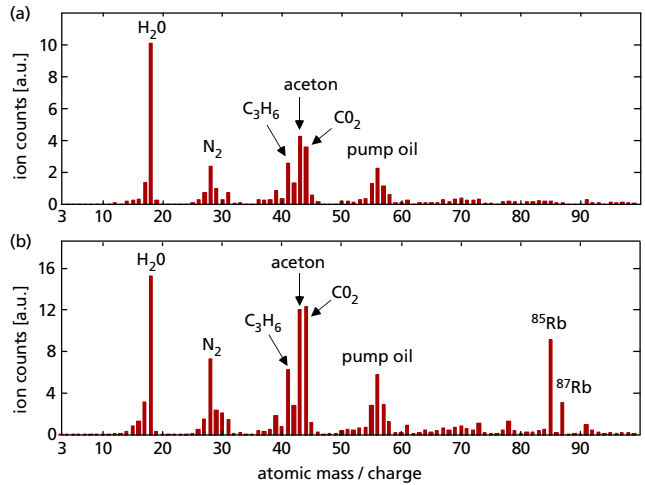}}}
\caption{(Color online) Mass spectra taken with a quadrupole mass spectrometer \cite{qpspec}. (a) Background spectrum at $\sim 10^{-8}$ mbar, before activating the dispenser. (b) Spectrum during dispenser operation at $\sim 10^{-7}$ mbar. The emission of the dispenser shows a clear signature of the rubidium isotopes $^{85}\mbox{Rb}$ and $^{87}\mbox{Rb}$ according to the natural isotope ratio.}
\label{fig:MassSpectrometer}
\end{figure}

We characterize the constituents emitted from a heated rubidium dispenser source by means of a quadrupole mass spectrometer \cite{qpspec}. The dispenser source is directed towards the mass spectrometer and is at ultrahigh vacuum ($\sim 10^{-8}$ mbar) in a chamber, which is pumped by a rotary vane pump and a turbo molecular pump, similar to the setup of the field ionization detector. Figure \ref{fig:MassSpectrometer}(a) shows the mass spectrum of the residual background gas in the chamber with the dispenser source turned off. The spectrum was taken across 96 atomic mass units and averaged over $40$s to reduce statistical noise. We identify remaining air constituents like water ($\mbox{H}_2 \mbox{O}$), nitrogen ($\mbox{N}_2$) and carbon dioxide ($\mbox{C} \mbox{O}_2$) as well as acetone remnants from cleaning solutions and components of pump oil.

\begin{figure}[tb]
\centerline{\scalebox{0.65}{\includegraphics{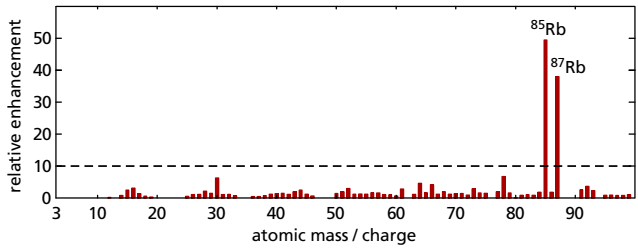}}}
\caption{(Color online) Mass spectrum during dispenser operation (Fig. \ref{fig:MassSpectrometer}(b)) normalized to the background spectrum (Fig. \ref{fig:MassSpectrometer}(a)). While the background signal increases by not more than a factor of three, the rubidium lines increases by a factor of up to 50.}
\label{fig:MassSpectrometer2}
\end{figure}

After activating the rubidium dispenser source at $5.9$A, the pressure in the chamber rises and reaches its steady state value at $\sim 10^{-7}$ mbar after a few minutes. Figure \ref{fig:MassSpectrometer}(b) shows the corresponding mass spectrum. As expected, this spectrum contains the signal of the rubidium isotopes $^{85}\mbox{Rb}$ and $^{87}\mbox{Rb}$ with a relative weight of $^{85}\mbox{Rb}/^{87}\mbox{Rb}=1/0.33$ close to the natural isotope ratio of $1/0.38$ \cite{lide2001}. In contrast to the background spectrum [Fig. \ref{fig:MassSpectrometer}(a)], we observe not only the appearance of rubidium lines but also a general increase in the signal strength of all background lines. We do not attribute this increased background to a direct emission from the dispenser source, but rather to an increased outgassing of vacuum components (dispenser shell, supply cables, electrical feedthrough) now at a higher temperature. To quantify the change in the spectrum due to the dispenser source, Fig. \ref{fig:MassSpectrometer2} shows the relative signal increase for each mass number compared to the background spectrum of Fig. \ref{fig:MassSpectrometer}(a). While the overall signal is increased not more than by a factor of three, the rubidium lines increased by a factor of up to 50. Thus rubidium is the element which spikes out of the background. We make use of this observation in the interpretation of the experimental data taken with the field ionization detector.

Using a quadrupole mass spectrometer we also have access to the ionization energy of each detected constituent. In the spectrometer, the neutral particles are first ionized by an electron beam and then the derivative ions are detected with respect to their mass to charge ratio. By changing the acceleration voltage of the electrons, we can set an upper threshold for the ionization energy of detected particles. At low acceleration voltages only the rubidium signal is present (ionization energy $4.18 \mbox{eV}$ \cite{Steck2008}). By increasing the electron's voltage the first background molecule to be detected is water with an ionization energy of $12.3 \mbox{eV}$. Thus all gas constituents in our vacuum chamber have ionization energies at least three times larger than rubidium. We conclude that in our setup rubidium is by far the most sensitive element to field ionization.



\begin{thebibliography}{34}
\bibitem[{ion()}]{ionizationrate}
\bibinfo{note}{According to Eq. 7, the ionisation rate drops dramatically when
  lowering the detector voltage. Proper ionization signals have been observed
  down to 500V with a maximum counting rate of about $20 \mbox{s}^{-1}$.}
\end{thebibliography}

\end{document}